\documentclass[grl]{agutex}

  \usepackage[dvips]{graphicx}
\authorrunninghead{DIMITRAKOUDIS ET AL.}

\titlerunninghead{STORM-TIME ULF WAVE RADIAL DIFFUSION}

\authoraddr{Corresponding author: S. Dimitrakoudis,
National Observatory of Athens, Institute for 
Astronomy, Astrophysics, Space Applications and Remote Sensing 
GR-15236, Penteli, Greece.
(sdimis@noa.gr)}

\begin{document}

%
%

\title{ Accurately specifying storm-time ULF wave radial
diffusion in the radiation belts }
%
%

%
%



 \authors{Stavros Dimitrakoudis,\altaffilmark{1} 
Ian R. Mann, \altaffilmark{2} Georgios Balasis, \altaffilmark{1}
 Constantinos Papadimitriou, \altaffilmark{1,3}
Anastasios Anastasiadis, \altaffilmark{1}
Ioannis A. Daglis, \altaffilmark{3,1}}

\altaffiltext{1}{National Observatory of Athens, Institute for 
Astronomy, Astrophysics, Space Applications and Remote Sensing 
GR-15236, Penteli, Greece.}
\altaffiltext{2}
{Department of Physics, University of Alberta, Edmonton, Alberta, Canada}
\altaffiltext{3}
{Section of Astrophysics, Astronomy and Mechanics, Department of Physics, 
University of Athens, Greece.}

%
%


\begin{abstract}
Ultra-low frequency (ULF) waves can contribute to the transport, acceleration and loss of electrons in the radiation belts through inward and outward diffusion. However, the most appropriate parameters to use to specify the ULF wave diffusion rates are unknown. Empirical representations of diffusion coefficients often use Kp; however, specifications using ULF wave power offer an improved physics-based approach. We use 11 years of ground-based magnetometer array measurements to statistically parameterise the ULF wave power with Kp, solar wind speed, solar wind dynamic pressure and Dst. We find Kp is the best single parameter to specify the statistical ULF wave power driving radial diffusion. Significantly, remarkable high energy tails exist in the ULF wave power distributions when expressed as a function of Dst. Two parameter ULF wave power specifications using Dst as well as Kp provide a better statistical representation of storm-time radial diffusion than any single variable alone.    

\end{abstract}

%
%

%

\begin{article}

%
%

\section{Introduction}
\label{sec1}

Ultra-low frequency (ULF) waves in the magnetosphere have long
been suggested as a likely factor affecting the 
acceleration and
diffusion of electrons in the outer radiation belt
\citep[e.g.][]{Falthammar1965,SchulzLanzerotti1974,Fei2006}. 
ULF wave power measurements on the ground or in-situ 
have led to various
attempts to derive an analytic expression for their 
effect on the
diffusion coefficient of particles, as a function of some geophysical 
index or solar wind parameter \citep[e.g.][]{BrautigamAlbert2000, 
Brautigam2005, Ozeke2012a, Ozeke2014}. 

The choice of which solar wind or geophysical parameter or parameters
should be used to most accurately specify the ULF wave power, and thereby the radial diffusion coefficients, is important but not yet well-understood. For example, the often used radial diffusion coefficient specification presented by \cite{BrautigamAlbert2000} parameterises the diffusion by Kp. Similarly, according to the approach of \cite{Brizard2001}, the radial diffusion coefficients are proportional to the power spectral density of ULF waves (see also \cite{Fei2006}) such that parameterisations of ULF wave power can be used to specify the radial diffusion coefficients.

In this regard, it has been shown for example that 
solar wind speed ($V_{sw}$) correlates with ULF power in the Pc3 to 5 
frequency range (1.7 to 100 mHz) 
\citep[e.g.][]{
Greenstadt1979, Mathie2000,  
Pahud2009, Simms2010, Rae2012},
with a likely physical mechanism being the generation of shear
flow instabilities along the magnetopause \citep{Cahill1992, Mann1999, Mathie2000}. 
Solar wind dynamic pressure changes ($P_{\mathrm{dyn}}$) have also been found to 
be correlated with ULF wave excitation 
\citep[e.g.][]{Kepko2002,
Takahashi2007, Kessel2008},
possibly due to the 
generation of compressional waves that transmit their
energy to field line resonances in the inner magnetosphere 
\citep{Kivelson1988, Lysak1992, Mann1995, Hartinger2011}. 
However, the nature of ULF wave excitation during magnetic storms is not fully understood. For example the morphology of the storm-time magnetosphere, as well as the solar wind structures which drive storm-time ULF waves such as interplanetary coronal mass ejections (CMEs) or fast solar wind streams/co-rotating interaction regions (CIRs), may be different from the conditions that prevail during nominal conditions. Consequently the ULF wave response of the system may be different during storm and non-storm times.

In this letter we present a study of 11 years of ground-based magnetometer observations of ULF waves by the IMAGE array, 
binned by four parameters: Kp, Dst,
solar wind dynamic pressure, $P_{\mathrm{dyn}}$, and solar wind speed, $V_{\mathrm{sw}}$. We firstly examine the distribution of the ULF wave power as a function of these parameters alone, to analyse which parameter might be most appropriate for the statistical representation of ULF wave power. The distributions of ULF wave power reveal that they have a very strong high energy tail. We hence further analyse the ULF wave power distribution as a function of multiple parameters, in particular to examine whether the storm-time disturbance index, Dst, should be used in addition to other parameters, such as Kp or $V_{\mathrm{sw}}$ which have more traditionally been used to parameterise ULF wave power and hence radial diffusion in the outer radiation belt.  

In section~\ref{sec2} we describe our dataset and processing method,
in section~\ref{sec3} we present our results, and we conclude in
section~4 with a summary and a short discussion.

\section{Data Processing}
\label{sec2}

\begin{table}
\label{deciles}
\caption{Decile borders for each binning parameter: Kp,
solar wind speed ($V_{\rm{sw}}$), solar wind dynamic pressure
 ($P_{\rm{dyn}}$) and Dst.}
\centering
\begin{tabular*}{\hsize}{@{\extracolsep{\fill}} l  c  c  c  c }
 \tableline  
 Decile border & Kp & $V_{\rm{sw}}$ (km/s)  & 
 $P_{\rm{dyn}}$ (nPa) & Dst (nT)\\
 \tableline 
 0 & 0 & 233 & 0.03 & 67  \\
 1 & 0.3 & 320 & 0.81 & 6  \\
 2 & 0.7 & 346 & 1.01 & 1  \\
 3 & 1 & 369 & 1.19 & -2  \\
 4 & 1.3 & 393 & 1.36 & -6  \\
 5 & 1.7 & 418 & 1.57 & -9  \\
 6 & 2 & 448 & 1.81 & -13  \\
 7 & 2.7 & 485 & 2.14 & -18  \\
 8 & 3 & 537 & 2.61 & -25 \\
 9 & 3.7 & 607 & 3.54 & -35  \\
10 &  9  &  1189  & 79.05  &  -422    \\
 \tableline 
\end{tabular*}
\end{table}

The power spectra used in this study are derived from 
11 years of
magnetic field measurements from nine stations in the IMAGE
(International Monitor for Auroral Geomagnetic Effects) 
 magnetometer array \citep{Tanskanen2009IMAGE}
(1 January 2000 to 31 December 2010): Uppsala, Nurmij\"arvi,
Domb\aa s, Ouluj\"arvi, R\o rvik, Sodankyl\"a, Kiruna,
Kevo, and Troms\o. 
The particular stations analysed were chosen on the basis of three conditions:
a) spanning a large range of geomagnetic latitude, b) spanning a small range of geomagnetic longitude, and c) $>95\%$ data coverage within our 
selected time period. 
We analysed measurements of the D (east) component of the 
magnetic field, 
sampled at 10 second cadence, and performed the continuous wavelet
transform with a Morlet mother function to derive hourly power spectra
\citep{Balasis2012,Balasis2013}.
The frequencies used were selected in the range from 0.6 to 19.85 mHz,
 spaced with a uniform linear step of
0.25 mHz (chosen for easy comparison with \citep{Ozeke2012a}, that used
fast Fourier Transform with a similar frequency step). These cover the Pc5 (up to 7 mHz) and Pc4 (7 to 20 mHz)
frequency ranges.
To remove most impacts from spectral features in the Pc4-5 ULF band which do not correspond to waves in the equatorial plane, such as nightside substorm bays, we only analysed data from daytime
hours from 0600 to 1800 MLT. We made use of the D-component
of the ground magnetic field measurements because it can be 
mapped to the azimuthal electric field in space assuming an Alfv\'{e}nic mode and a 90$^{\circ}$ rotation through the ionosphere, which is 
then used
to calculate the electric term of the radial diffusion coefficient
\citep{Ozeke2012a}. The mapping procedure we used 
\citep{Ozeke2009} assumes a dipole geometry for the magnetic field,
which is problematic as an approximation for the nightside
of the magnetosphere; but this is somewhat mitigated by our 
exclusive use of daytime measurements.

The ULF wave power measurements were then binned with Kp, $V_{\rm{sw}}$,
 $P_{\rm{dyn}}$ and Dst using data from the OMNI database \citep{King2005} at hourly resolution. 
In order to robustly compare the relative efficacy 
of each binning parameter, we divided the dataset into ten equal
deciles, ensuring that each bin for each different 
driving parameter has the
same number of data points. Table~1 shows the lowest
and highest values of the parameters during the 
period of observation, as well as the values that
denote the borders of their respective deciles. 
The total number of days is 4013, 
since there is a five-day gap 
in ULF measurement data from early 2000, which means we have
48216 daylight hours of data in total, and up to 4816 hourly spectra in 
each decile. 

Since the OMNI solar wind database is not complete for this interval, each $V_{\rm{sw}}$ decile has only 4809 data points, while
each $P_{\rm{dyn}}$ decile has 4768 data points. The difference in bin
sizes, with the ones for $V_{\rm{sw}}$ being $0.14\%$ smaller and the
ones for $P_{\rm{dyn}}$ being $1\%$ smaller than the ones for Kp and Dst,
can be considered to be statistically negligible. 
The total number of daylight hours for each station is further
affected by small data gaps interspersed throughout the 11 years,
with up to $5\%$ fewer hours in the case of R\o rvik station.  Such gaps 
have no effect on the relative size of the decile bins for each station,
and assuming the data gaps are randomly distributed they will only have a 
very small effect on the statistical comparisons from station to station. 

As shown below, the data demonstrate that the ULF wave power distributions have long high power tails, including for Dst. In the case of 
Dst we further calculate the highest ten percentiles, splitting the 
highest Dst decile further into 10 percentile bins in order to examine the storm-time variations of ULF wave power. Finally, we also 
derive two-dimensional probability distributions, to both examine the utility of using two parameter ULF wave specifications and to look for differences between ULF wave power during storm and 
non-storm times.

\section{Dependence of ULF wave Power on the Selected Parameters}
\label{sec3}

\begin{figure*}[h]
\noindent\includegraphics[width=40pc]{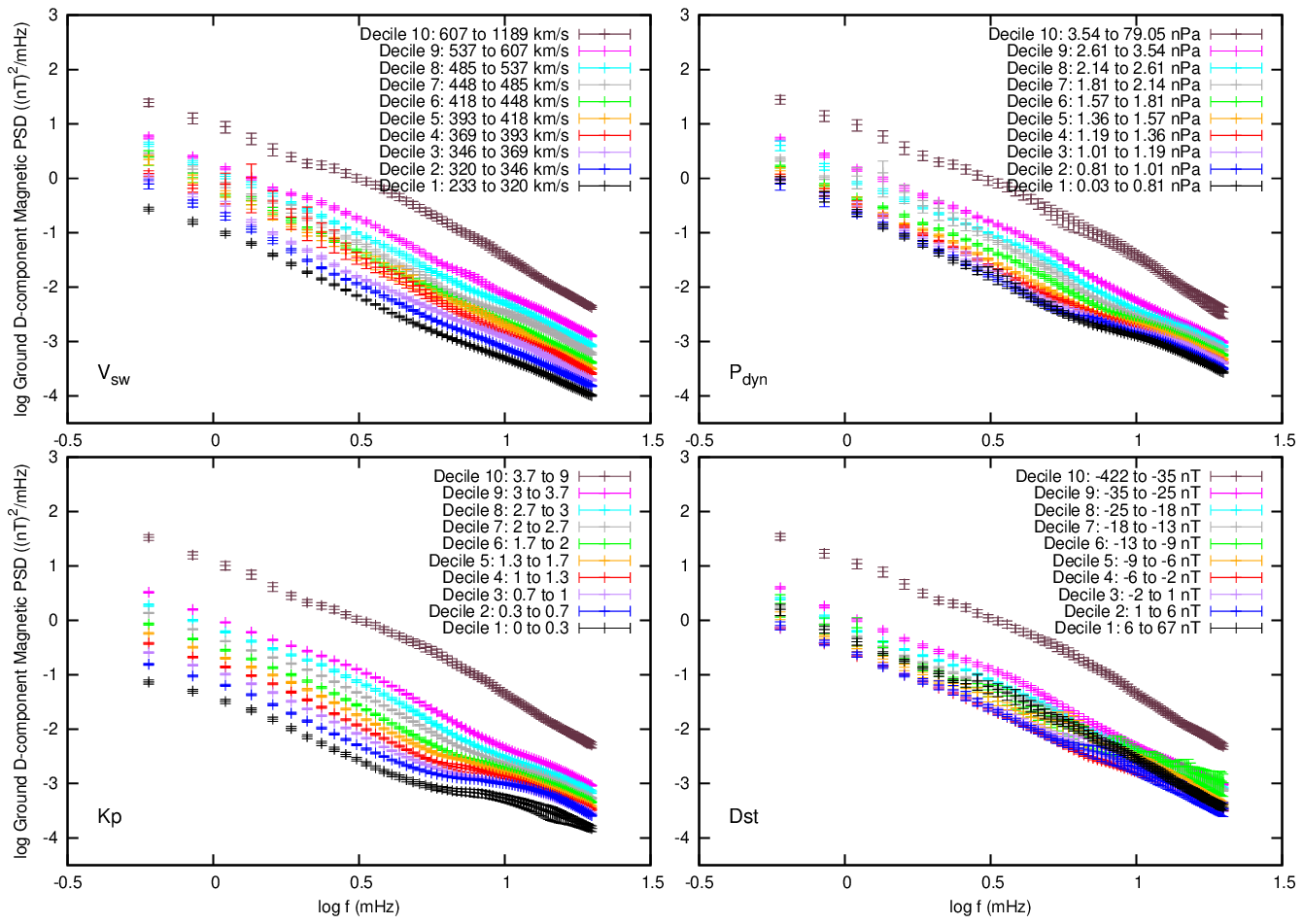}
\caption{
A comparison of mean magnetic field D-component 
power spectral densities as measured at Nurmij\"arvi station  (L=3.4), 
when binned by deciles, with Kp, solar wind speed, solar wind pressure, 
and Dst. The error bars show the standard error.
}
\label{parameters_4_NUR}
\end{figure*}

\begin{figure*}[h]
\noindent\includegraphics[width=40pc]{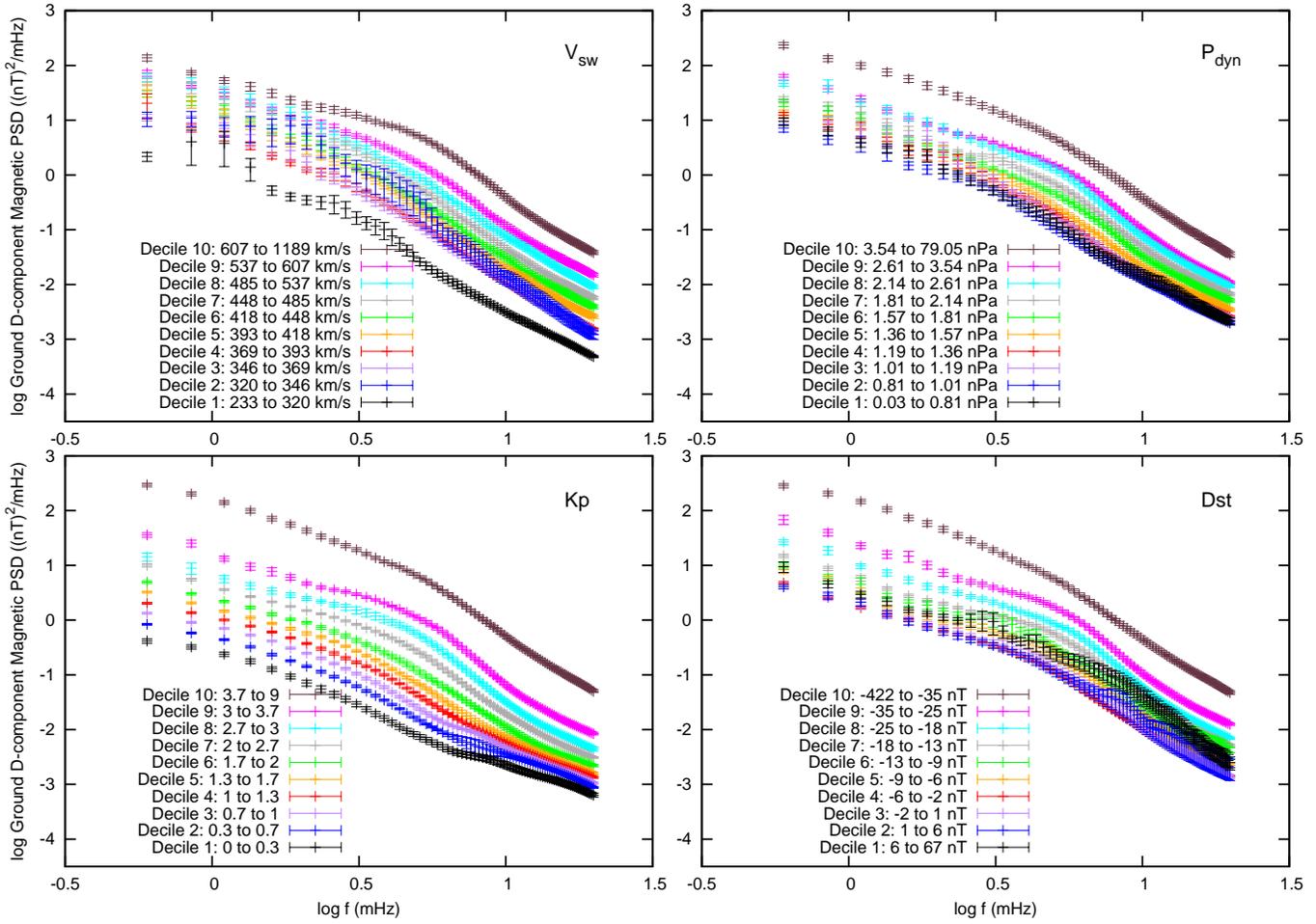}
\caption{As in Figure~\ref{parameters_4_NUR}
but for Troms\o \ station (L=6.46).
}
\label{parameters_4_TRO}
\end{figure*}

Figure~\ref{parameters_4_NUR} shows the mean PSDs and 
their standard errors in each decile for
Nurmij\"arvi (NUR) station at $L=3.4$ (geomagnetic
longitude $102.18^{\circ}$),
for all four parameters used in this study, while 
Figure~\ref{parameters_4_TRO} shows the same parameters for the Troms\o \ (TRO) station at $L=6.46$ (geomagnetic
longitude $102.9^{\circ}$). These stations span from mid-latitude to close to geosynchronous orbit, with L-values which span the outer radiation belt.
The following traits stand out (which also apply
to the observations from the intermediate latitude stations and which are not shown):

\begin{enumerate}
\item Binning by Kp provides the largest span in mean power, with smallest standard error in each decile, particularly in the 
Pc5 frequency range, and hence represents the best single parameter with which to specify radial diffusion. Binning by $V_{\mathrm{sw}}$ maintains a good span and has utility for radial diffusion simulations as well; the deciles of $P_{\mathrm{dyn}}$ and Dst do not provide good discriminators of ULF wave power,– however the ULF power in the tenth Dst decile is examined further below.

\item The tenth decile of the Dst distributions is clearly separated from the others, such that the mean of the 1-9 deciles are very similar but ULF wave power increases significantly once Dst $ < -35$ nT. $P_{\mathrm{dyn}}$ shows similar behavior; indeed the high energy tail of the ULF wave power distribution can also be seen in the Kp and the $V_{\rm{sw}}$ distributions as well. Perhaps more significantly, the separation of the tenth Dst decile from the remainder of the distribution is more pronounced at the lower-L station suggesting the impact of the storm-time dependence of ULF wave power might be relatively larger towards the inner edge of the outer belt.
\item There is also evidence, especially at the high L station, for a local enhancement in the frequency spectrum of the ULF wave power above a simple power law, which most likely indicates the impact of ULF wave power accumulation at the local field line resonance \cite[cf.][]{Rae2012}.
\end{enumerate} 

\begin{figure*}[h]
\noindent\includegraphics[width=40pc]
{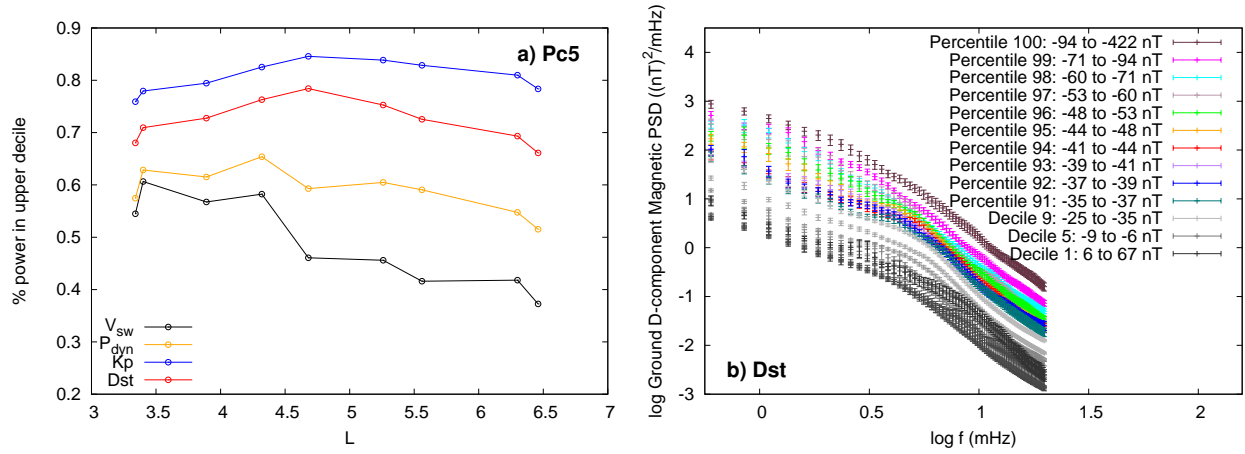}
\caption{a) Percentage of total power in the upper decile, for the Pc5
frequency range, for all four binning parameters and for all stations.
b) Mean magnetic field D-component 
power spectral densities as measured at Troms\o \ station (L=6.46),
binned by Dst as in Figure~\ref{parameters_4_TRO}, except that the top
decile is further split into percentiles (denoted by 91 - 100). 
The lower nine deciles are plotted in grayscale (only three of them, deciles 1, 5, and 9, appear in the legend, as indicative of the colour progression).
}
\label{rel_power_Pc5}
\end{figure*}

This L-dependence of the ULF power is explored in more detail in panel (a) of Figure~\ref{rel_power_Pc5},
where we
have integrated the power spectral density profile for each decile across the
Pc5 frequency range, and derived the ratio $Y_i$ of
the power carried by the upper decile to the total power, 
where $i$ refers to the binning parameter. 
In the Pc5 frequency range, for Kp and Dst we
see a change in $Y_i$ at around $L = 4.7$, and 
for $P_{\mathrm{dyn}}$ and $V_{\mathrm{sw}}$ at around $L = 4.3$,
which nevertheless 
have the same effect: a decrease of $Y_i$ with L in the outer magnetosphere.
This emphasises the strength of the high power tail, especially at low L.
This behaviour 
likely reflects on the physical mechanisms behind the generation of
the Pc4-5 ULF waves, however, a detailed discussion of this is beyond the scope of the current study. Nonetheless, it shows that the ULF wave power distributions are dominated by a high power tail. 

Panel (b) of Figure~\ref{rel_power_Pc5} examines the nature of the high ULF wave power tail by further splitting the top decile into 10 percentiles. What is incredibly clear is that not only are magnetic storms with large Dst magnitude associated with very large ULF wave power, but also that the ULF wave power defined on the basis of an 11 year data base remains well ordered by the Dst parameter even at the percent level in this high power tail. Overall, this suggests that studies of the impacts of ULF waves in the radiation belts during magnetic storms should additionally include the Dst index together with another parameter, such as Kp, when specifying the ULF wave diffusion rates. Such an approach may provide an improved ULF wave power specification for the purposes of modelling radial diffusion. 

\begin{figure*}[h]
\noindent\includegraphics[width=40pc]
{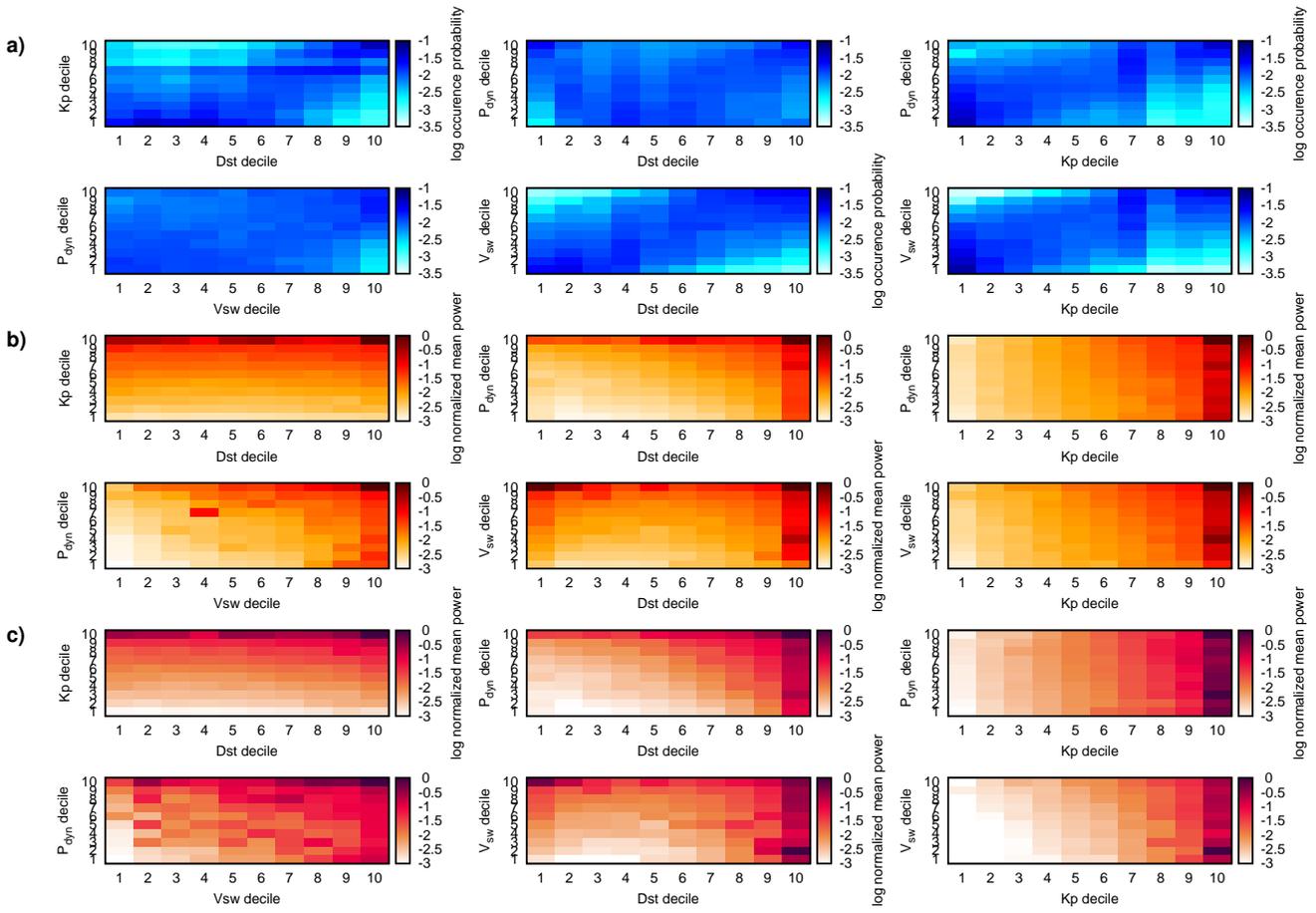}
\caption{a) Occurrence probability distributions for
pairs of deciles of combinations of the parameters Kp, $V_{\rm{sw}}$, $P_{\rm{dyn}}$ and Dst. 
b) Mean ULF wave power for each pair of deciles, as in
panel (a), for Nurmij\"arvi (NUR) station.
c) Same as panel (b) but for Troms\o \ (TRO) station.
}
\label{heat_maps}
\end{figure*}

Figure~\ref{heat_maps} shows occurrence probability distribution (panel(a)) as well as the mean ULF wave power in each bin as a function of pairs of deciles of two of the parameters Kp, $V_{\rm{sw}}$,
 $P_{\rm{dyn}}$ and Dst for the NUR (panel(b)) and TRO (panel (c)) stations. Panel (a) provides an indication of the parts of the distribution with reliable statistics, 
and in panels (b) and (c) the mean power is normalised to 
the bin with maximum mean power. 
Panels (b) and (c) show that the range of mean ULF wave power is very large, spanning 3 orders of magnitude. As can be clearly seen in this Figure, Kp is an excellent discriminator of ULF wave power, with some evidence that conditions of higher $V_{\rm{sw}}$ and $P_{\rm{dyn}}$ also contribute additionally to higher ULF wave power for given Kp. It is also very clear that more negative Dst also contributes to higher ULF power, especially at the lower L station. Recall that the highest value of Dst in the 10th Dst decile has the value of $-35$ nT (cf. Table 1) such that effectively all storm times are captured in this decile (cf. also the percent level discrimination of the ULF wave power as a function of Dst shown in panel (b) of Figure 3). Significantly, this indicates that the storm time magnetosphere on average has significantly elevated ULF wave power for the same activity index, e.g., Kp, as compared to non-storm times.

 We have also verified that the 
data analysis presented here using data from the IMAGE magnetometer array and the Morlet wavelet to characterise the ULF wave power reproduces the electric field diffusion coefficients reported in \cite{Ozeke2012a} and which were derived using data largely from the Canadian CARISMA array
\citep{Mann2008CARISMA} and a Fast Fourier Transform approach. The details will be reported in a more extensive publication elsewhere.

\section{Summary and Discussion}
\label{sec4}

We have presented a new statistical analysis of the 
connection between ULF wave activity in the magnetosphere and 
four solar wind and geomagnetic index driving parameters (Kp, solar wind speed, solar wind dynamic pressure and Dst) using ground-based magnetometer observations of ULF wave power spanning one solar 
cycle. 
Unlike a similar study undertaken by \cite{Huang2010}, which 
evaluated the effect of various parameters on ULF waves in geostationary
orbit, our focus on the D-component of multiple ground magnetometer measurements allowed us to probe the power going into the electric
term of radial diffusion, at L-shells covering the full width of the
outer radiation belt. 
Although \cite{Huang2010} focused on magnetic field measurements
with a linear binning method, which does not allow for casual 
comparisons with our results even when those are for an L-shell
corresponding to geosynchronous orbit, it is worth noting that they 
also observed a correlation of the increase of ULF power with increases
of Kp, Dst, and $V_{\mathrm{sw}}$.
Since the ULF wave electric field diffusion usually dominates over the magnetic field diffusion \citep[e.g.,][]{ Ozeke2012a, Tu2012} then our results have significant implications for the most appropriate specification of the ULF wave power, and hence the radial diffusion rates in the outer electron radiation belt.  
We conclude that of these single parameters Kp provides the best capability to represent the dynamic range of ULF wave power, which is driven in the magnetosphere. Furthermore, it becomes apparent that 
the ULF wave power present during magnetic storms (as characterised by Dst, which measures the average magnetic depression near the equator as a result of the ring current formed by such storms) is not the same as that averaged over the entire solar cycle. We show conclusively that the level of ULF wave power present in the magnetosphere is larger at storm times than at non-storm times for the same level of driving conditions, at least as specified by Kp (which measures the average global field disturbances), solar wind speed,  and solar wind dynamic pressure drivers. As a result, we conclude that including Dst as an additional parameter provides an improved method for specifying the ULF wave power and hence the rate of ULF-wave radial diffusion in the radiation belts. We suggest that future studies should examine ULF wave power characterisations which use both Kp and Dst to specify the power and, as such, could be compared to studies which use the observed ULF wave power in individual storms to drive diffusion. Future studies could additionally examine the ULF wave power relationship to whether the magnitude of Dst is increasing or decreasing - thereby examining the statistical differences between ULF wave excitation in the main and recovery phases of magnetic storms. 

It has been shown in the past
\citep[e.g.][]{Lotoaniu2006, Mann2012book} 
that ULF wave power penetration to low L is
connected with decreases in Dst. Here that is seen with greater
clarity, since at low L the ULF wave power
comes predominantly from the upper decile of Dst measurements, which 
corresponds to Dst $< -35$ nT, while at high L 
the rest of the deciles have a larger contribution to the
overall dynamic range of ULF wave power. Significantly, since ground PSD measurements of ULF waves
can be directly translated to the electric term of the radial
diffusion coefficient on the equatorial plane in space 
\citep[e.g.][]{Ozeke2012a}, our results have implications for the most appropriate parameters to use for characterising ULF wave driven radial diffusion. 
Accurate specification of such diffusive transport rates, especially at the inner edge of the outer radiation belts, will likely be crucial for developing a physical understanding of the nature of the penetration of ULF wave power to low -L \citep[cf.][]{Lotoaniu2006}. Such analyses may also shed light on the physical processes responsible for the reported correlations between the location of this inner boundary and the plasmapause \citep[e.g.,][]{li2006correlation} (itself correlated with Dst see, e.g., \cite[][]{o2003empirical}), as well as the recent reports of the apparently largely impenetrable nature of the inner edge of the outer belt \cite[][]{baker2014impenetrable}.


%
%
%
%
%
%
%

\begin{acknowledgments}
This work has received funding
from the European Union's Seventh Framework
Programme (FP7-SPACE-2011-1) under grant agreement n.
284520 for the MAARBLE (Monitoring, Analyzing and Assessing
Radiation Belt Loss and Energization) collaborative research
project.
We acknowledge support from the ``Hellenic
National Space Weather Research Network'' co-financed by
the European Union (European Social Fund – ESF) and Greek
national funds through the Operational Program ``Education and
Lifelong Learning'' of the National Strategic Reference Framework
(NSRF) – Research Funding Program ``Thales. Investing in
knowledge society through the European Social Fund''.
We also acknowledge use of NASA/GSFC's Space Physics Data Facility's 
OMNIWeb (or CDAWeb or ftp) service, and OMNI data.
We thank the institutes who maintain the IMAGE Magnetometer Array. IRM was supported by A Discovery Grant from Canadian NSERC.
\end{acknowledgments}

%
%
%
%
%
%
%
%
%

%
%
\end{article}
\end{document}